\documentclass[a4paper,12pt,nofootinbib]{revtex4}
\usepackage{bm}
\usepackage{amsmath}
\usepackage{amsfonts}

\begin{document}
\title{Norm-overlap formula of Hartree-Fock-Bogoliubov states
with odd number parity}
\author{Makito Oi}
\affiliation{Institute of Natural Sciences, Senshu University, 
3-8-1 Kanda-Jinbocho, Chiyoda-ku,Tokyo 101-0051, Japan}
\email{m.oi@isc.senshu-u.ac.jp}

\author{Takahiro Mizusaki}
\affiliation{Institute of Natural Sciences, Senshu University, 
3-8-1 Kanda-Jinbocho, Chiyoda-ku,Tokyo 101-0051, Japan}
\date{\today}

\begin{abstract}
  A formula to calculate a norm overlap between 
  Hartree-Fock-Bogoliubov (HFB) states with the
  odd number parity (one quasi-particle excited states) 
  is derived with help of the Grassmann numbers 
  and the Fermion coherent states.
  The final form of the formula is expressed 
  in terms of a product of 
  the Pfaffian for a neighboring even-even system (the zero
  quasi-particle state), and an extra factor consisting of the
  Bogoliubov transformation matrix and the anti-symmetric matrix in
  Thouless' HFB ansatz for the even-even system.
\end{abstract}
\maketitle

\section{Introduction}
The Hartree-Fock-Bogoliubov (HFB) method has been a powerful method
in descriptions of nuclear states \cite{RS80}. The reason is that the method
can deal with the two most important correlations in interacting 
many-body nuclear systems, that is, deformation and pairing.
These correlations are effectively taken into account by breaking
relevant symmetries (the rotational and gauge symmetries).
 As a consequence,  however, the conservation laws of angular momentum and
the particle numbers are violated.

Restorations of these broken symmetries are achieved through quantum
number projections, but there have been difficulties to overcome
in carrying out practical calculations of the projection. The major
problem lies in the calculation of the so-called norm overlap kernels,
which is necessary in the projections.
The evaluation of the norm overlap is particularly difficult 
in angular momentum projection because the rotational symmetry is 
associated with the non-Abelian SO(3) group.

An analytical formula was demonstrated to express the norm overlap 
by Onishi and Yoshida \cite{OY66}.
But due to a square root operation appearing in the formula,
the relative sign (or phase) of HFB states at various points 
in the Euler space needs to be determined 
with respect to the reference HFB state, with an additional effort.

Hara, Hayashi and Ring \cite{HHR82} were the first to perform 
a numerical calculation of angular momentum projection.
They made use of the continuity and differentiability of the overlap,
in order to determine the sign.

However, it was later found that the assignment was sometimes 
very hard to be achieved due to a peculiar nature of the overlap.
Such a case was seen in the cranked HFB wave functions, and
it was discovered that the so-called ``nodal lines'' (a collection of 
zeros of the overlap) are the source of the problem \cite{OT05}.
A method to overcome this problem was presented in Ref.\cite{OT05}, 
and an improvement to the method was recently found by the authors \cite{OM11}.
With this method based on the Onishi formula, the sign problem was solved.

Robledo proposed a totally different approach to the sign problem \cite{Rob09}.
Making use of the Grassmann numbers and the Fermion coherent state,
he was successful to remove the square root operation in the norm overlap 
formula. His new formula is described by means of the Pfaffian.
His approach to rely on the Grassmann algebra 
is not only mathematically elegant, but also quite powerful 
in practical computations of overlaps of many-body operators.
There can be many applications to be discovered through his new methodology.

In this paper, we would like to present such an application:
a formula to evaluate a norm overlap between two HFB states
with the odd number parity, 
which corresponds to nuclei with the odd-mass number.

\section{Hartree-Fock-Bogoliubov states}
When the total number of constituent particles is even, 
the corresponding HFB ansatz (Thouless ansatz) is given as
\begin{equation}
  |\text{HFB}\rangle = \mathcal{N}
  \exp\left(\sum_{i<j}^MZ_{ij}c^{\dag}_ic_{j}^{\dag}\right)|0\rangle.
\end{equation}
Although this ansatz breaks the particle number conservation,
the number parity is kept to be positive, 
or $(-1)^{2n}$ ($n$ integer) \cite{RS80}.
In the above expression, 
the dimension of the configuration space is given as $M$, and
the (true) vacuum state $|0\rangle$ is defined as
\begin{equation}
  c_i|0\rangle = 0,
\end{equation}
where the single-particle annihilation (creation) operator of the state $i$ 
is expressed as $c_i$ ($c_i^{\dag}$). 

Quasi-particle bases are introduced through a canonical transformation
called the Bogoliubov transformation $\mathcal{W}$, which is 
\begin{equation}
  \left(\begin{array}{c} c \\ c^{\dag}\end{array}\right)
    = \mathcal{W}
  \left(\begin{array}{c} \beta \\ \beta^{\dag}\end{array}\right),
\end{equation}
where
\begin{equation}
  \mathcal{W} =\left(\begin{array}{cc} U & V^*\\ V & U^*\end{array}\right).
\end{equation}
Here, $\beta_i^{\dag}$ and $\beta_i$ are
creation and annihilation operators for quasi-particles, respectively.
The $U$ and $V$ are both $M\times M$ matrices, and can be regarded as the 
variational parameters in the HFB theory. These matrices satisfy the
properties coming from the unitarity of the Bogoliubov transformation,
that is, $\mathcal{W}^{\dag}\mathcal{W} = \mathcal{W}\mathcal{W}^{\dag} = \mathcal{I}$ ($\mathcal{I}$ is the identity matrix) \cite{RS80}.

The anti-symmetric matrix $Z$ appearing in the HFB ansatz is related to the
$UV$ matrices as
\begin{equation}
  Z = (VU^{-1})^*.
  \label{ZVU}
\end{equation}
The anti-symmetry of the above matrix can be easily confirmed 
by the properties possessed by the $UV$ matrices.

The normalization constant $\mathcal{N}$ is calculated to be
\begin{equation}
  \mathcal{N} = \sqrt{\text{det}U},
\end{equation}
by using the Onishi formula \cite{OH80}.
In many calculations of physical interest, 
such as one-dimensional cranked HFB states, 
$U$ is a real matrix, so that $\mathcal{N}$ is a real number.

The HFB state $|\Phi\rangle$ is the vacuum of the quasi-particles,
that is, the following condition is satisfied,
\begin{equation}
  \beta_k|\Phi\rangle = 0.
\end{equation}
In the quasi-particle bases, the corresponding energy spectrum is given as
\begin{equation}
  \hat{H}_{\text{HFB}} = E_0 + \sum_k E_k \beta_k^{\dag}\beta_k.
\end{equation}

Excited states with many quasi-particles can be produced by
operating the quasi-particle creation operators to the HFB vacuum. 
For example, a one-quasi-particle excited state is expressed 
in the framework of the HFB theory as \cite{RS80},
\begin{equation}
  |\Phi_k\rangle = \beta_k^{\dag}|\Phi\rangle.
  \label{HFB_odd}
\end{equation}
The excited energy corresponding to this state is $E_k$.
 As explained in p.250 of Ref.\cite{RS80}, the new state $|\Phi_k\rangle$
has the negative number parity, which means that the state $|\Phi_k\rangle$
 corresponds to an odd-mass nucleus, that is,
a neighbor to the even-even nucleus described by $|\Phi\rangle$.

For the sake of simplicity, the isospin degree of freedom 
is not considered in the present paper,
but the extension of the theory can be easily made.

\section{Pfaffian formula to  norm overlap}
\subsection{Case of even number parity}
The case of the even number parity was well studied by Robledo \cite{Rob09}, 
and the Pfaffian formula of norm overlap kernels 
was derived for even-even nuclear systems
for the first time in his work.
But his conventions for the mathematical objects are slightly different
from Ref.\cite{NO85}, which is employed in the present work. 
For the sake of consistency in the following discussions, 
it will be convenient to derive the formula again here 
with our mathematical conventions.

The essential point to derive the Pfaffian formula is to introduce
the Fermion coherent state and its completeness.
The Fermion coherent state \cite{NO85} reads
\begin{equation}
  |\bm{\xi}\rangle = \text{e}^{-\sum_{i}\xi_{i}c_{i}^{\dag}}|0\rangle,
\end{equation}
and it satisfies by definition the eigenvalue equation,
\begin{equation}
  c_i|\bm{\xi}\rangle = \xi_i|\bm{\xi}\rangle,
  \label{eigen-eq}
\end{equation}
where $\xi_i$ represents the Grassmann number.
The Grassmann numbers follow the anticommutation rule, that is,
$\xi_i\xi_j + \xi_j\xi_i = 0$. In the special case of $i=j$, there holds
$\xi_i^2=0.$

The completeness is given as
\begin{equation}
  \int\prod_{i}d\xi_i^*d\xi_i \exp\left(-\sum_j\xi_j^*\xi_j\right)
  |\bm{\xi}\rangle\langle\bm{\xi}| = 1.
\end{equation}

Let us write a HFB state as the following:
\begin{equation}
  |\Phi^{(p)}\rangle = \hat{T}(Z^{(p)},c^{\dag})|0\rangle,
  \label{Thouless}
\end{equation}
where an operator $\hat{T}$ is introduced as
\begin{equation}
  \hat{T}(Z,c^{\dag}) = 
\exp\left(\frac{1}{2}\sum_{ij}Z_{ij}c_i^{\dag}c_j^{\dag}\right),
\end{equation}
for $p=0,1$.
In this notation, $\hat{T}(Z,c^{\dag})^{\dag}=\hat{T}(-Z^*,c)$, because
$Z$ is an anti-symmetric matrix.

The overlap between two HFB states is thus expressed in the following way.
\begin{eqnarray}
&&  \langle\Phi^{(0)}|\Phi^{(1)}\rangle \label{overlap} \\ \nonumber
&=&\langle 0|\hat{T}(-Z^{(0)*},c)\hat{T}(Z^{(1)},c^{\dag})|0\rangle \\ \nonumber
&=&\int\prod_{\alpha}d\xi_{\alpha}^*d\xi_{\alpha}
  \text{e}^{-\sum_{\beta}\xi_{\beta}^*\beta_{\beta}}
  \langle 0|\hat{T}(-Z^{(0)*},c)|\bm{\xi}\rangle\langle\bm{\xi}|\hat{T}(Z^{(1)},c^{\dag})|0\rangle \\ \nonumber
&=&\int\prod_{\alpha}d\xi_{\alpha}^*d\xi_{\alpha}
  \text{e}^{-\sum_{\beta}\xi_{\beta}^*\beta_{\beta}}
  {T}(-Z^{(0)*},\xi)
  {T}(Z^{(1)},\xi^{*}).
\end{eqnarray}
In the last line, the operator $\hat{T}$ is replaced with a Grassmann-number
quantity $T$ due to Eq.(\ref{eigen-eq}), which is 
\begin{equation}
  T(Z,\xi^*) = \exp\left(\frac{1}{2}\sum_{ij}Z_{ij}\xi_i^*\xi_j^*\right).
\end{equation}
In addition, a property of
\begin{equation}
  \langle 0|\bm{\xi}\rangle = 1
\end{equation}
is used in the last line.

Let us write the integrand in Eq.(\ref{overlap}) as $G(\bar{\xi})$,
that is, 
\begin{equation}
  G(\bar{\xi}) = 
  \text{e}^{-\sum_{\beta}\xi_{\beta}^*\beta_{\beta}}
  {T}(-Z^{(0)*},\xi)
  {T}(Z^{(1)},\xi^{*})
\end{equation}
As demonstrated by Robledo \cite{Rob09},
$G$ can be summarized to be a Gaussian with the Grassmann number:
\begin{equation}
  {G}(\bar{\xi}) = 
  \exp\left(\frac{1}{2}\bar{\zeta}^t\mathbb{Z}
    \bar{\zeta}\right),
  \label{Gauss}
\end{equation}
where Grassmann vectors in his work are defined as
\begin{equation}
  \bar{\zeta}^t \equiv (\xi_1^*,\xi_2^*,\cdots,\xi_M^*,
  \xi_1,\xi_2,\cdots,\xi_M),
  \label{vector-robledo}
\end{equation}
and a matrix $\mathbb{Z}$ with the $2M$ dimension is equal to
\begin{equation}
  \mathbb{Z} = \left(\begin{array}{cc} Z^{(1)} & -\mathcal{I}\\
      \mathcal{I} & -Z^{(0)*}\end{array}\right).
\end{equation}
$\mathcal{I}$ is the $M\times M$ identity matrix.
Apparently, $\mathbb{Z}$ is anti-symmetric.

In the present work, the following ordering for Grassmann vectors is employed,
\begin{equation}
  \bar{\xi}^t \equiv (\xi_1^*,\xi_2^*,\cdots,\xi_M^*,
  \xi_M,\xi_{M-1},\cdots,\xi_1),
  \label{vector}
\end{equation}
which is the same as in Ref.\cite{NO85}.
The transformation from $\bar{\zeta}$ to $\bar{\xi}$ is achieved as
\begin{equation}
  \bar{\xi} = L\bar{\zeta},
\end{equation}
where a linear transformation is given by
\begin{equation}
  L = \left( \begin{array}{cc}\mathcal{I} & 0\\ 0 & \Lambda \end{array}\right).
\end{equation}
A $M\times M$ matrix $\Lambda$ is defined as
\begin{equation}
  \Lambda_{ij} = \delta_{i+j,M+1}.
 \end{equation}
This matrix satisfies $L=L^t = L^{-1}$ ($\Lambda=\Lambda^t = \Lambda^{-1}$),  
and
\begin{equation}
\text{det}(L) = \text{det}(\Lambda)=(-1)^{M(M-1)/2}.
\end{equation}
After the transformation, the Gaussian $G$ is rewritten as
\begin{equation}
  G(\bar{\zeta}) = G(\bar{\xi})=\exp\left(\frac{1}{2}\bar{\xi}^t
  \mathbb{X}\bar{\xi}\right).
\end{equation}
The relation between $\mathbb{Z}$ and $\mathbb{X}$ is 
$  \mathbb{X}=L^t\mathbb{Z}L$, so that
\begin{equation}
  \mathbb{X} = \left(\begin{array}{cc}
      Z^{(1)} & -\Lambda \\
      \Lambda & -\Lambda Z^{(0)*}\Lambda
    \end{array}\right).
  \label{Xmatrix}
\end{equation}

Thanks to a mathematical theorem \cite{Zum62,BBF00}, it is always possible
to find a decomposition of an anti-symmetric matrix into a matrix product
\begin{equation}
  \mathbb{X} = R^{t}\mathbb{J}R,
  \label{decomposition}
\end{equation}
where $R$ is a regular matrix ($\exists R^{-1}$) and 
$\mathbb{J}$ corresponds to a canonical form of $\mathbb{X}$, that is, 
\begin{equation}
  \mathbb{J} = \left( \begin{array}{cc} \mathcal{O} & \mathcal{I} \\ 
      -\mathcal{I} & \mathcal{O}\end{array}\right).
\end{equation}

The new Grassmann bases $\bar{\eta}$ associated with the canonical form 
is obtained by a linear transformation of the original bases $\bar{\xi}$,
\begin{equation}
  \bar{\eta} = R\bar{\xi}.
\end{equation}
For the sake of convenience in subsequent discussions, 
let us write the inverse transformation as the following
\begin{equation}
  \bar{\xi}
  =
  R^{-1}\bar{\eta}
  \Longrightarrow
  \left(\begin{array}{c}
   \xi^* \\ \xi
    \end{array}\right) 
  =
  \left(\begin{array}{cc}
    \mathcal{R}_{11} & \mathcal{R}_{12}\\
    \mathcal{R}_{21} & \mathcal{R}_{22}
    \end{array}\right)
  \left(\begin{array}{c}
   \eta^* \\ \eta
    \end{array}\right).
\end{equation}
It should be noted that the index of $\eta$ and $\xi$ runs in a reverse order, 
as in Eq.(\ref{vector}). 

In the new bases $\bar{\eta}$, $G$ has an expression of
\begin{eqnarray}
  \label{R}
  G(\bar{\xi}) & \rightarrow & 
  G(\bar{\eta}) = \exp\left(\frac{1}{2}\bar{\eta}^t
    (R^{-1})^t\mathbb{X}R^{-1}\bar{\eta}\right)    \\ \nonumber
  &=& \exp(\frac{1}{2}\bar{\eta}^t\mathbb{J}\bar{\eta})
  = \exp\left(\sum_{\alpha}^M\eta^*_{\alpha}\eta_{M+1-\alpha}\right).
\end{eqnarray}
Let us introduce a notation $\bar{\alpha}$ here for convenience in subsequent
discussions, which is defined as
\begin{equation}
  \bar{\alpha} \equiv M+1-\alpha.
\end{equation}
Using the property of the Grassmann number ($\eta^2_i=0$), a Taylor expansion
of the exponential can be greatly simplified to a sum of bilinear polynomials.
It is thus possible to write $G$ as
\begin{equation}
G(\bar{\eta}) = \prod_{\alpha}^M(1+\eta_{\alpha}^*\eta_{\bar{\alpha}}).
  \label{Rproduct}
\end{equation}
Furthermore, it is important to understand 
that only the non-vanishing contribution of such polynomials
in the $2M$-dimensional Grassmann integral comes from the integrand
in which all the bilinear pairs $(\eta_{\alpha}^*,\eta_{\alpha})$ appear.
After an expansion of the above product, such paired term appear 
in the form of $\prod_{\alpha}\eta_{\alpha}^*\eta_{\bar{\alpha}}$.
To convert this expression to the standard paired expression, 
there is a useful identity
\begin{equation}
  \prod_{\alpha}^M\eta_{\alpha}^*\eta_{\bar{\alpha}} 
  = (-)^{M/2}\prod_{\alpha}^M\eta_{\alpha}^*\eta_{\alpha},
\end{equation}
where $M$ is an even integer.

With this identity, the integral of the paired polynomial is simplified,
and the calculation is easily done.
\begin{equation}
  \int \prod_{\alpha}^Md\eta_{\alpha}^* d\eta_{\alpha}
  \prod_{\beta}^M\eta^*_{\beta}\eta_{\beta}
  = (-1)^{M}.
  \label{nonvanish}
\end{equation}
The integral of the other polynomials give a null contribution 
to the overlap calculation of our interest.

Noting that the Jacobian for the bases transformation $R$ is $\text{det}(R)$,
the integral of $G(\bar{\xi})$, or the overlap, becomes
\begin{eqnarray}
  \langle\Phi^{(0)}|\Phi^{(1)}\rangle
  &=& \int\prod_{\alpha}d\xi_{\alpha}d\xi_{\alpha}G(\bar{\xi}) \\
  \nonumber
  &=& (-1)^{M/2}\text{det}(R)\int\prod_{\alpha}d\eta^*_{\alpha}d\eta_{\alpha}
  \prod_{\beta}\eta_{\beta}^*\eta_{\beta} \\ \nonumber
  &=& (-1)^{M/2} (-1)^M \text{det}(R).
\end{eqnarray}

Using another identities, that is, 
\begin{equation}
  \text{Pf}(\mathbb{X}) = \text{det}(R)\text{Pf}(\mathbb{J}),
\end{equation}
and 
\begin{equation}
  \text{Pf}(\mathbb{J}) = (-1)^{M(M-1)/2},
\end{equation}
the final expression is obtained as
\begin{equation}
  \langle\Phi^{(0)}|\Phi^{(1)}\rangle = (-1)^{M(M+2)/2}\text{Pf}(\mathbb{X})
  =\text{Pf}(\mathbb{X}).
  \label{pfaffian}
\end{equation}
The phase factor gives rise to $(-1)^{M(M+2)/2}=+1$ for even $M$,
and it is different from Robledo's formula,
which is $(-1)^{M(M+1)/2}$.
This is because of the difference in the definitions of the Grassmann vectors,
Eqs. (\ref{vector-robledo}) and (\ref{vector}), and it is simply explained as
$\text{Pf}(\mathbb{X}) = \det(L)\text{Pf}(\mathbb{Z})
= (-1)^{M(M+1)/2}\text{Pf}(\mathbb{Z})$ \footnote{$(-1)^{M(M+1)/2} = (-1)^{M(M-1)/2}$ because $(-1)^M = 1$ for $M$ being even.}.

\subsection{Case of odd number parity}
Next, let us consider two HFB states with the odd number parity.
Following Eq.(\ref{HFB_odd}), they are expressed as
\begin{equation}
  |\Phi^{(0)}_k\rangle = \beta^{\dag(0)}_k|\Phi^{(0)}\rangle; \quad
  |\Phi^{(1)}_{k'}\rangle = \beta^{\dag(1)}_{k'}|\Phi^{(1)}\rangle.
\end{equation}
Here, the HFB states $|\Phi^{(p)}\rangle \ (p=0,1)$ are considered
to be states with the even number parity that are given by Eq.(\ref{Thouless}).

Below, a formula is derived for an overlap between the two states 
with the odd number parity, that is,
\begin{equation}
  \langle\Phi_k^{(0)}|\Phi_{k'}^{(1)}\rangle
  =  \langle\Phi^{(0)}|\beta_{k}^{(0)}\beta_{k'}^{\dag(1)}|\Phi^{(1)}\rangle,
  \label{odd-overlap}
\end{equation}
which is expressed in terms of the Bogoliubov transformation matrices 
($U$ and $V$) and the inverse of $\mathbb{X}$ given in Eq.(\ref{Xmatrix}).

It may be worth noting that a re-arrangement of a product 
$\beta_{k}^{(0)}\beta_{k'}^{\dag(1)} = -\beta_{k'}^{\dag(1)}\beta_{k}^{(0)}
+ (U^{\dag(0)}U^{(1)} + V^{\dag(0)}V^{(1)})_{kk'}$ 
does not help very much to simplify the overlap because the vacuum condition is
applied only to the associated annihilation operator. In other words,
$\beta_k^{(p)}|\Phi^{(q)}\rangle = 0$ only when $p=q$.

In the single-particle bases, the product $\beta_{k}^{(0)}\beta_{k'}^{\dag(1)}$
is expanded as
\begin{equation}
  \beta_{k}^{(0)}\beta_{k'}^{\dag(1)} = 
  \hat{\mathfrak{S}}_{kk'}(c,c^{\dag}) +  \hat{\mathfrak{K}}_{kk'}(c,c^{\dag}).
\end{equation}
Two operators $\hat{\mathfrak{S}}$ and $\hat{\mathfrak{K}}$
are bilinear functions of the single-particle creation and annihilation
operators, which are defined as
\begin{eqnarray}
  \hat{\mathfrak{S}}_{kk'} &=& \sum_{ij}\left(
    \mathcal{S}^{kk'}_{ij}c_ic_{j}^{\dag}
    + \mathcal{T}^{kk'}_{ij}c_{i}^{\dag}c_{j}\right), \\
  \hat{\mathfrak{K}}_{kk'} &=& \sum_{ij}\left(
    \mathcal{K}^{kk'}_{ij}c_ic_{j}
    + \mathcal{L}^{kk'}_{ij}c_{i}^{\dag}c_{j}^{\dag}\right).    
\end{eqnarray}
Let us call $\hat{\mathfrak{S}}$ the normal operator
while $\hat{\mathfrak{K}}$ the dangerous operator.
The matrix elements in the right hand side 
are given in terms of the $UV$ matrices.
\begin{equation}
 \mathbb{W}^{kk'} \equiv \left(\begin{array}{cc}
      \mathcal{L}^{kk'}_{ij} & \mathcal{S}^{kk'}_{ij} \\
      \mathcal{T}^{kk'}_{ij} & \mathcal{K}^{kk'}_{ij} 
    \end{array}\right)
  =
  \left(\begin{array}{cc}
      (V^{\dag})_{ki}^{(0)}U_{jk'}^{(1)} &
      (U^{\dag})_{ki}^{(0)}U_{jk'}^{(1)} \\
      (V^{\dag})_{ki}^{(0)}V_{jk'}^{(1)} &
      (U^{\dag})_{ki}^{(0)}V_{jk'}^{(1)} \\  
    \end{array}\right)
\end{equation}
The calculation of the overlap given in Eq.(\ref{odd-overlap}) is
thus reduced to a sum of the overlaps of the normal and dangerous operators,  
with respect to the HFB states with the even number parity, that is,
\begin{equation}
  \langle\Phi^{(0)}_{k}|\Phi_{k'}^{(1)}\rangle
  = \langle\Phi^{(0)}|\hat{\mathfrak{S}}_{kk'}|\Phi^{(1)}\rangle
  + \langle\Phi^{(0)}|\hat{\mathfrak{K}}_{kk'}|\Phi^{(1)}\rangle.
\end{equation}
In the following subsections, each term in the right-hand side
is considered separately.

\subsection{The normal operator $\hat{\mathfrak{S}}$}
The essential ingredient of the calculation here is an evaluation of
$\langle\Phi^{(0)}|c_{l}c_{l'}^{\dag}|\Phi^{(1)}\rangle$.
We begin with an insertion of the completeness for the Fermion coherent 
state between the creation and annihilation operators as
\begin{eqnarray}
  &&\langle\Phi^{(0)}|c_{i}c_{j}^{\dag}|\Phi^{(1)}\rangle \\ \nonumber
  &=& \int \prod_{\alpha}^Md\xi_{\alpha}^*d\xi_{\alpha}
  \text{e}^{-\sum_{\beta}\xi_{\beta}^*\xi_{\beta}} 
  \langle\Phi^{(0)}|c_i|\bm{\xi}\rangle
  \langle\bm{\xi}|c_{j}^{\dag}|\Phi^{(1)}\rangle \\ \nonumber
  &=& \int \prod_{\alpha}^Md\xi_{\alpha}^*d\xi_{\alpha}
  \text{e}^{-\sum_{\beta}\xi_{\beta}^*\xi_{\beta}} 
  \langle\Phi^{(0)}|\bm{\xi}\rangle
  \langle\bm{\xi}|\Phi^{(1)}\rangle \xi_i\xi_{j}^* \\
  &\equiv & \int \prod_{\alpha}^Md\xi_{\alpha}^*d\xi_{\alpha}
  G(\bar{\xi})\xi_i\xi_{j}^*.
  \label{Eq21}
\end{eqnarray}
The $G(\bar{\xi})$ has the common structure seen in the even 
number parity case, that is, Eq.(\ref{Gauss}).

Considering that $\xi_{i}\xi_j^*$ gives rise to three kinds of terms
consisting of bilinear expressions of $\eta_{\alpha}$ and $\eta_{\alpha}^*$,
what we need to calculate in Eq.(\ref{Eq21}) 
are 
$G(\bar{\eta})\eta_i\eta_j$, 
$G(\bar{\eta})\eta_i^*\eta_j^*$,
and $G(\bar{\eta})\eta_i^*\eta_j$.
However, from a simple analysis, the first two cases go to zero
after integrations. 
It is thus enough to consider the last case.

Noting the reverse order in the index for $\eta$ and $\xi$,
we have
\begin{eqnarray}
  \xi_i^* &=& \sum_{j=1}^M\left(\mathcal{R}_{11}\right)_{ij}\eta_j^* + 
  \left(\mathcal{R}_{12}\right)_{ij}\eta_{\bar{j}}, 
    \label{Transf1}\\
  \xi_i &=& \sum_{j=1}^M\left(\mathcal{R}_{21}\right)_{ij}\eta_j^* + 
  \left(\mathcal{R}_{22}\right)_{ij}\eta_{\bar{j}}.
  \label{Transf2}
\end{eqnarray}
From the property of the Grassmann integral, 
which is given in Eq.(\ref{nonvanish}), 
the non-vanishing contribution comes from the term of
$\eta_{\bar{j}}\eta_{j'}^*$.
Due to an identity relation shown in Eq.(\ref{nonvanish}),
there must hold a relation between the indices $j$ and $j'$,
which is to be explained below.

Let us consider a product 
$(1+\eta_{j'}^*\eta_{\bar{j}'})(1+\eta_{\bar{j'}}^*\eta_{j'})(1+\eta_{j}^*\eta_{\bar{j}})(1+\eta_{\bar{j}}^*\eta_{j})\eta_{\bar{j}}\eta_{j'}^*$, assuming that $j\ne j'$.
The first four factors are always included in a representation of 
$G(\bar{\eta})$, given in Eq.(\ref{Rproduct}). 
Not only the factors commute mutually, 
but also with the other factors in Eq.(\ref{Rproduct}).
Then, a re-ordering of the product simplifies the first and third factors, 
thanks to a property of the Grassmann number, that is, 
$(1+\eta_{j'}^*\eta_{\bar{j}'})\eta_{j'}^*(1+\eta_{\bar{j'}}^*\eta_{j'})(1+\eta_{j}^*\eta_{\bar{j}})\eta_{\bar{j}}(1+\eta_{\bar{j}}^*\eta_{j}) = \eta_{j'}^*(1+\eta_{\bar{j'}}^*\eta_{j'})\eta_{\bar{j}}(1+\eta_{\bar{j}}^*\eta_{j})$
Then, an expansion of the product gives rise to
 $\eta_{j'}^*\eta_{\bar{j}} - \eta_{j'}^*\eta_j\eta_{\bar{j}}^*\eta_{\bar{j}}
- \eta_{j'}^*\eta_{j'}\eta_{\bar{j'}}^*\eta_{\bar{j}} 
+ \eta_{j'}^*\eta_{j'}\eta_{\bar{j}}^*\eta_{\bar{j}}\eta_{\bar{j'}}^*\eta_j$. 
All these four terms give a null contribution to the integral 
due to a presence of unpaired Grassmann numbers.
In order to maintain the pair structure as seen in Eq.(\ref{nonvanish}),
there must hold $j=j'$. Then, the relevant product becomes
$(1+\eta_{j}^*\eta_{\bar{j}})(1+\eta_{\bar{j}}^*\eta_{j})\eta_{j}^*\eta_{\bar{j}}
=\eta_{j}^*\eta_{\bar{j}}(1+\eta_{\bar{j}}^*\eta_{j})
=-\eta_{\bar{j}}^*\eta_{\bar{j}}\eta_{j}^*\eta_{j}$, which does not vanish
after integration.
The result is thus summarized in an identity
\begin{equation}
  \int \prod_{\alpha}^Md\eta_{\alpha}^*d\eta_{\alpha}G(\bar{\eta})\eta_k^*
  \eta_{\bar{k'}} = (-1)^{3M/2}\delta_{kk'}.
  \label{res_integl}
\end{equation}

Noting that the Jacobian is given by $\text{det}(R)$ and putting all the 
above result together, the whole integral Eq.(\ref{Eq21}) becomes
\begin{eqnarray}
&&\int\prod_{\alpha} 
d\xi_{\alpha}^*d\xi_{\alpha}G(\bar{\xi})\xi_{i}\xi_{j}^* \\ \nonumber
&=&
\text{det}(R)\int\prod_{\alpha} 
d\eta_{\alpha}^*d\eta_{\alpha}G(\bar{\eta}) 
\sum_{\beta\gamma}\left(
  \eta_{\beta}^*\eta_{\bar{\gamma}}
  \left(\mathcal{R}_{21}\right)_{i\beta}\left(\mathcal{R}_{12}\right)_{j\gamma}
  \right.
  \\ \nonumber
  &&\left. -\eta_{\gamma}^*\eta_{\bar{\beta}}
  \left(\mathcal{R}_{22}\right)_{i\beta}\left(\mathcal{R}_{11}\right)_{j\gamma}
  \right),
\\ \nonumber
&=&
(-1)^{3M/2}\text{det}(R)
  \left(\mathcal{R}_{21}\mathcal{R}_{12}^t
    -\mathcal{R}_{22}\mathcal{R}_{11}^t\right)_{ij}
\\ \nonumber
&=& \text{Pf}(\mathbb{X})   \left(\mathcal{R}_{21}\mathcal{R}_{12}^t
    -\mathcal{R}_{22}\mathcal{R}_{11}^t\right)_{ij}
\end{eqnarray}
In the last line, we have used relations $\text{Pf}(\mathbb{X}) = \text{det}(R) \ \text{Pf}(\mathbb{J})$ and $\text{Pf}(\mathbb{J}) =(-1)^{M(M-1)/2}$.
Also, a fact that $M(M+2)/2$ is an even integer was used.

The similar calculation is performed for $\langle\Phi^{(0)}|c_i^{\dag}c_{j}|\Phi^{(1)}\rangle$. In this case, it is important to exchange the order of a product
of the creation and annihilation operators, that is, 
$c_i^{\dag}c_j = -c_jc_i^{\dag}+\delta_{ij}$, for the convenience in applying
the completeness of the Fermion coherent state. The first term corresponds to
the exactly same result as obtained above, except the sign and the change
in the indices ($i\leftrightarrow j$). Whereas, the second term is a C-number,
 so that the overlap is simply proportional to 
$\langle\Phi^{(0)}|\Phi^{(1)}\rangle.$

The final expression for the normal operator becomes the following
\begin{equation}
 \langle\Phi^{(0}|\hat{\mathfrak{S}}_{kk'}|\Phi^{(1)}\rangle 
  = \langle\Phi^{(0)}|\Phi^{(1)}\rangle  \text{Tr}\left(T^{kk'}\mathfrak{N}_1 
    + S^{kk'}\mathfrak{N}_2\right),
\end{equation}
where 
\begin{eqnarray}
  \mathfrak{N}_1 &=& \mathcal{R}_{12}\mathcal{R}_{21}^t
  - \mathcal{R}_{11}\mathcal{R}_{22}^t, \\
  \mathfrak{N}_2 &=& \mathcal{R}_{22}\mathcal{R}_{11}^t 
  - \mathcal{R}_{21}\mathcal{R}_{12}^t
   + \mathcal{I}.
\end{eqnarray}

In obtaining the above expression, the result obtained in Eq.(\ref{pfaffian})
is also used.

\subsection{The dangerous operators $\mathfrak{K}$}
The essential ingredients in this subsection is overlaps of the so-called
``dangerous terms'' in the HFB theory, which are expressed as
$\langle\Phi^{(0)}|c^{\dag}_ic^{\dag}_j|\Phi^{(1)}\rangle$ and
$\langle\Phi^{(0)}|c_ic_j|\Phi^{(1)}\rangle$.
Because a complex conjugate of one type of the dangerous terms
corresponds to the other, it is sufficient to consider a mathematical 
analysis for one of the two terms. Let us take the first type here.
That is, 
\begin{eqnarray}
  &&\langle\Phi^{(0)}|c_{i}^{\dag}c_{j}^{\dag}|\Phi^{(1)}\rangle \\ \nonumber
  &=& \int \prod_{\alpha}^Md\xi_{\alpha}^*d\xi_{\alpha}
  \text{e}^{-\sum_{\alpha}\xi_{\alpha}^*\xi_{\alpha}} 
  \langle\Phi^{(0)}|\bm{\xi}\rangle\langle\bm{\xi}|c_i^{\dag}c_{j}^{\dag}
  |\Phi^{(1)}\rangle \\ \nonumber
  &=& \int \prod_{\alpha}^Md\xi_{\alpha}^*d\xi_{\alpha}
  \text{e}^{-\sum_{\alpha}\xi_{\alpha}^*\xi_{\alpha}} 
  \langle\Phi^{(0)}|\bm{\xi}\rangle
  \langle\bm{\xi}|\Phi^{(1)}\rangle \xi_i^*\xi_{j}^* \\
  &\equiv & \int \prod_{\alpha}^Md\xi_{\alpha}^*d\xi_{\alpha}
  G(\bar{\xi})\xi_i^*\xi_{j}^*.
  \label{Eq55}
\end{eqnarray}
According to Eqs.(\ref{Transf1}) and (\ref{Transf2}),
a product $\xi_i^*\xi_j^*$ is expanded in terms of bilinear polynomials
of $\eta$ and $\eta^*$. As discussed in the previous section, only the type
of terms $\eta_k^*\eta_{\bar{k'}}$ contributes to the integral if $k=k'$.
The result is given in Eq.(\ref{res_integl}).
The final result is thus obtained as
\begin{equation}
 \langle\Phi^{(0)}|c_{i}^{\dag}c_{j}^{\dag}|\Phi^{(1)}\rangle   
 =\text{Pf}(\mathbb{X}) \left(\mathcal{R}_{11}\mathcal{R}_{12}^t 
-\mathcal{R}_{12}\mathcal{R}_{11}^t\right)_{ij}.
\end{equation}

The other dangerous term can be obtained in a similar way,
\begin{equation}
 \langle\Phi^{(0)}|c_{i}c_{j}|\Phi^{(1)}\rangle   
 =\text{Pf}(\mathbb{X}) \left(\mathcal{R}_{21}\mathcal{R}_{22}^t 
-\mathcal{R}_{22}\mathcal{R}_{21}^t\right)_{ij}.
\end{equation}

By combining these results, the dangerous part becomes
\begin{equation}
 \langle\Phi^{(0}|\hat{\mathfrak{K}}_{kk'}|\Phi^{(1)}\rangle 
  = \langle\Phi^{(0)}|\Phi^{(1)}\rangle  \text{Tr}\left(
    \mathcal{K}^{kk'}\mathfrak{D}_1 
    + \mathcal{L}^{kk'}\mathfrak{D}_2\right),
\end{equation}
where
\begin{eqnarray}
  \mathfrak{D}_1 &=& \mathcal{R}_{22}\mathcal{R}_{21}^t 
  - \mathcal{R}_{21}\mathcal{R}_{22}^t, \\
  \mathfrak{D}_2 &=& \mathcal{R}_{12}\mathcal{R}_{11}^t
  - \mathcal{R}_{11}\mathcal{R}_{12}^t.
\end{eqnarray}

\subsection{The final expression}
The sum of the overlaps of the normal and dangerous operators
gives rise to the final form of the overlap formula
in the case of the odd number parity.

Before writing down the formula, however, 
it is worth considering one more thing here.
In the above discussion, we introduced the quantities expressed in terms of 
 the inverse of the transformation matrix $R$,
that is, $\mathfrak{N}_{i}$ and $\mathfrak{D}_i$ $(i=1,2)$.
It is possible to express these quantities directly 
through the inverse of $\mathbb{X}$. The demonstration can be shown
simply by taking the inverse of the both sides of Eq.(\ref{decomposition}).
The result is
\begin{eqnarray}
  \mathbb{X}^{-1} &=& -R^{-1}\mathbb{J}(R^{-1})^t \\
  &=&
  \left(\begin{array}{cc}
    \mathfrak{D}_{2} & \mathfrak{N}_{1}\\
    \mathfrak{N}_{2}-\mathcal{I} & \mathfrak{D}_{1}
    \end{array}\right).
\end{eqnarray}

The formula is therefore obtained as
\begin{equation}
\langle\Phi^{(0)}_k|\Phi^{(1)}_{k'}\rangle
= \langle\Phi^{(0)}|\Phi^{(1)}\rangle 
\text{Tr}(\mathbb{W}^{kk'}\mathbb{X}^{-1}+\mathcal{S}^{kk'}).
\label{formula}
\end{equation}
The final form of the formula is independent of the bases transformation
matrix $R$.

The advantage of this formula is that the overlap of the
odd-number-parity is expressed in terms of the quantities
obtained for the neighboring even-even nucleus $|\Phi^{(i)}\rangle$:
the overlap of the even-number-parity, i.e., $\text{Pf}(\mathbb{X})$,
and the Bogoliubov transformation $\mathcal{W}$.
Although the inverse of the matrix $\mathbb{X}$ needs to be 
computed for the formula, the matrix itself can be expressed
through the quantities calculated for the even-even system 
(See.Eq(\ref{Xmatrix})).
In other words, the quantum number projections can be
done simultaneously for an even-even nucleus 
and the neighboring odd system, without significant efforts.

The similar procedure can be applied
to a derivation of formulae for multiple quasi-particle excited states
(with more than one quasi-particle).
The basic structure is a product of the Pfaffian of a certain even-even system
and the factors representing each quasi-particles expressed 
in terms of  $\mathcal{W}$, including the inverse of $\mathbb{X}$.

\section{Summary}
A formula Eq.(\ref{formula}) was demonstrated
so as to calculate a norm overlap for the HFB states with the
negative number parity (one quasi-particle states), 
which correspond to odd-mass nuclei.
The Grassmann algebra and the Fermion coherent state are employed,
so as to allow the Pfaffian to describe the overlap.

The formula has a factorized structure, which consists of the norm overlap
for an even-even system and part described in terms of the 
Bogoliubov transformation matrix,
as well as the inverse of the matrix given in Eq.(\ref{Xmatrix}).
This structure is beneficial because both of the systems
with the positive and negative number parities can be studied 
at the same time by means of angular momentum projection.

Recently, Avez and Bender presented the similar work \cite{AB11}, as
well as Bertsch and Robledo \cite{BR11}. 
These works, including ours, result in the Pfaffian, 
which were initially demonstrated by Robledo.
However, in their HFB wave functions, fully blocked unpaired particles 
are assumed
($V=1, U=0$ in terms of the Bogoliubov transformation), 
which are different from our ansatz Eq.(\ref{HFB_odd}) 
based on the $UV$ exchange approximation for
multiple quasi-particle excited states. In addition, the mathematical 
representation of the final results are significantly different from 
each other.

\begin{acknowledgments}
The authors thank for useful discussions with Professor R. C. Johnson.
This work is financially supported with a research grant in Senshu University.
\end{acknowledgments}


\begin{thebibliography}{99}
  \bibitem{RS80} P. Ring and P. Schuck, Nuclear Many-Body Problem, 
    Springer-Verlarg, 1980.
  \bibitem{OY66} N. Onishi and S. Yoshida, Nucl. Phys. 80, 367 (1966).
  \bibitem{HHR82} K. Hara, Y. Hayashi, P. Ring, Nucl. Phys. A358, 14 (1982).
  \bibitem{OT05}M. Oi and N. Tajima, Phys. Lett. B 606, 43 (2005).
    \bibitem{OM11} M. Oi and T. Mizusaki, in preparation.
  \bibitem{Rob09}L. M. Robledo, Phys. Rev. C. 79, 021302R (2009).
  \bibitem{OH80} N. Onishi and T. Horibata, Prog. Theor. Phys. 64, 1650 (1980).
  \bibitem{NO85} J. W. Negele and H. Orland, {\it Quantum Theory of Finite
      Systems} (MIT Press, Cambridge, MA/London, 1985).
  \bibitem{Zum62} B. Zumino, J. Math. Phys. 3, 1055 (1962).
  \bibitem{BBF00} P. Benner, R. Byers, H. Fassbender, V. Mehrmann, D. Watkins,
    Electronic Transactions on Numerical Analysis 11, 85 (2000).
  \bibitem{AB11} B. Avez and M. Bender, arXiv[nucl-th]:1109.2078v1 (2011).
  \bibitem{BR11} G.F. Bertsch and L. M. Robledo, arXiv[nucl-th]:1108.5479v1 
    (2011).
\end{thebibliography}
\end{document}